# Bichromatic field generation from double-four-wave mixing in a double-electromagnetically induced transparency system

**Yang Liu, Jinghui Wu, Dongsheng Ding, Baosen Shi and Guangcan Guo**

Key Laboratory of Quantum Information, Chinese Academy of Sciences,

University of Science and Technology of China,

Hefei, 230026

**Abstract.** We demonstrate the double electromagnetically induced transparency (double-EIT) and double four-wave mixing (double-FWM) based on a new scheme of non-degenerate four-wave mixing (FWM) involving five levels of a cold $^{85}$Rb atomic ensemble, in which the double-EIT windows are used to transmit the probe field and enhance the third-order nonlinear susceptibility. The phase-matching conditions for both four-wave mixings could be satisfied simultaneously. The frequency of one component of the generated bichromatic field is less than the other by the ground-state hyperfine splitting (3GHz). This specially designed experimental scheme to simultaneously generate different nonlinear wave-mixing processes is expected to find applications in quantum information processing and cross phase modulation. Our results agree well with the theoretical simulation.

**Contents**

## 1. Introduction

Electromagnetically induced transparency (EIT) is a quantum interference phenomenon that allows for the transmission of light through an otherwise opaque atomic medium [1]. The importance of EIT stems from the fact that it results in a greatly enhanced linear and nonlinear susceptibility in the medium near the induced transparent frequency window and is associated with steep dispersion. So EIT is widely used in light slowing and light storage [8-11], biphoton wave packet generation and modulation in cold atoms [12-15], cross phase modulation [16-19] and so on. Under EIT conditions, the four-wave mixing (FWM) process can be resonantly enhanced [4-7], two or more wave-mixing processes (for example, FWM and six-wave mixing) could coexist and interfere with each other [21, 22]. In Ref. [23], the authors realized the simultaneous control of two four-wave-mixing fields by changing the atomic density, the intensity, and the detuning of the coupling field in a triple-Λ-type atomic system in a hot $^{85}$Rb cell. In their recent work of [24], they also successfully observed the generation of four and six strongly correlated and anticorrelated laser fields from such a system.

The bichromatic EIT in cold atoms has been experimentally studied [28], in which the multiple-peaked probe absorption spectra under bichromatic coupling fields have been observed. The FWM signal with multi-peaks structure in such a system has been performed [29]. Bichromatic optical solitons could be generated in a cold atomic cloud in the condition of a pulsed probe field and a pulsed FWM field of considerably different frequency [30, 31]. In this paper, we report a new experimental study of the double-FWM in a five-level atomic system. Two phase-matching conditions of FWM could be satisfied simultaneously and the generated two FWM signals are observed. As compared to the similar work in Ref. [23, 24], our work shows that such an experimental scheme could generate a so-called bichromatic field. That is, the generated two FWM signals have not only controllable relative strength but also the same spatial and temporal mode and polarization. Such a special third-order nonlinear wave mixing may have potential applications in quantum information process and cross phase modulation at low light level.

## 2. Theoretical model

Fig.1 shows the scheme of our experiment. The States |1> and |3> are the degenerate Zeeman sublevels corresponding respectively to the magnetic quantum numbers $m_F=-3$ and $m_F=-1$ of the ground-state hyper fine level F =3 of $^{85}$Rb. The state |2> corresponds to the other hyper fine level $|5^2S_{1/2}F=2, m_F=-1>$. And |4> and |5> are the excited states: $|5^2P_{1/2}F=3, m_F=-2>$ and $|5^2P_{3/2} F=3, m_F=-3>$, respectively. The coupling field $\omega_{c1}$ with a frequency detuning $\Delta_1$ couples the transition $|2>\rightarrow|4>$, and similarly the coupling field $\omega_{c2}$ with a detuning $\Delta_s$ couples the transition $|3>\rightarrow|4>$. The two coupling beams are spatially mode matched by coupling them into the same single mode fiber and collimated to a beam diameter of 3.2 mm in the vacuum chamber. Both of the two coupling beams are σ- polarized. The pump laser $\omega_p$ is σ+ polarized and red-detuned $\Delta_p$ from the transition $|1>\rightarrow|5>$. The probe field $\omega_s$ has the same frequency but opposite circular polarization with the coupling beam $\omega_{c2}$. The presence of the pump beam, two coupling beams and the probe beam will construct two possible FWM paths ($|3>\rightarrow|4>\rightarrow|1>\rightarrow|5>\rightarrow|3>$ and $|2>\rightarrow|4>\rightarrow|1>\rightarrow|5>\rightarrow|2>$). There are two possible phase-matching

conditions $\vec{k}_p + \vec{k}_{c1} + \vec{k}_s + \vec{k}_{as1} = 0$ and $\vec{k}_p + \vec{k}_{c2} + \vec{k}_s + \vec{k}_{as2} = 0$, where $\vec{k}_{as1}$ and $\vec{k}_{as2}$ are the wave vectors of two generated anti-Stokes fields from the two FWM processes respectively.

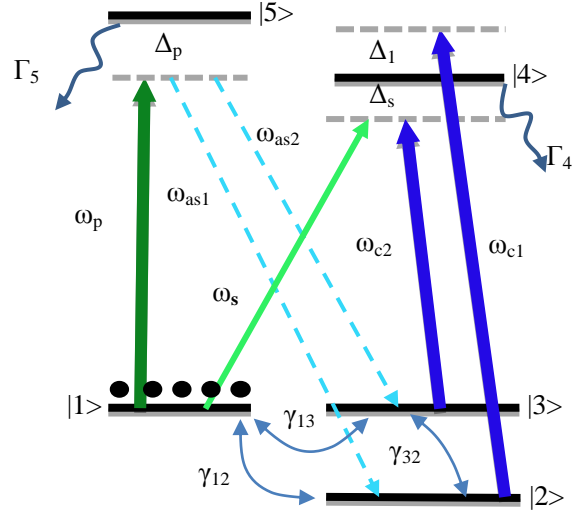

**Figure 1.** (color online) Schematic of the bichromatic field generation. $\Gamma_4=2\pi\times5.75$MHz and $\Gamma_5==2\pi\times6.07$MHz are the decay rates from the states $|4\rangle$ and $|5\rangle$ respectively. $\gamma_{12}$, $\gamma_{13}$ and $\gamma_{32}$ are the decoherences between states $|1\rangle$ and $|2\rangle$, states $|1\rangle$ and $|3\rangle$, and states $|3\rangle$ and $|2\rangle$, respectively.

In the interaction picture, the Hamiltonian of this five-level system could be written as:
$$H_{\text{int}} = \hbar(\Delta_p|5\rangle\langle5| - (\Delta_1 - \Delta_s)|2\rangle\langle2| + \Delta_s|3\rangle\langle3|$$
$$+ \Delta_s|4\rangle\langle4|) - \hbar(\Omega_p|5\rangle\langle1| + \Omega'_{as1}|5\rangle\langle2| + \Omega'_{as2}|5\rangle\langle3| \quad (1)$$
$$+ \Omega_s|4\rangle\langle1| + \Omega_{c1}|4\rangle\langle2| + \Omega_{c2}|4\rangle\langle3| + H.c)$$

Where $\Omega_i$ are the Rabi frequency of beam $i$, $\Omega_{as1}'=\Omega_{as1}\exp[i(\Delta_s-\Delta_1)t]$ and $\Omega_{as2}'=\Omega_{as2}\exp[i\Delta_s t]$ (which come from the rotating wave approximation). The master equation for the atomic density operator will determine the dynamics of the laser-driven atomic system.

$$\frac{d\rho}{dt} = \frac{1}{i\hbar}[H_{\text{int}},\rho] + \sum_{m=1,2,3}\frac{\Gamma_{4m}}{2}[2\hat{\sigma}_{m4}\rho\hat{\sigma}_{4m} - \hat{\sigma}_{44}\rho - \rho\hat{\sigma}_{44}]$$
$$+ \sum_{n=1,2,3}\frac{\Gamma_{5n}}{2}[2\hat{\sigma}_{n5}\rho\hat{\sigma}_{5n} - \hat{\sigma}_{55}\rho - \rho\hat{\sigma}_{55}] \quad (2)$$
$$+ \sum_{i=2,3,4,5}\frac{\gamma_{ideph}}{2}[2\hat{\sigma}_{ii}\rho\hat{\sigma}_{ii} - \hat{\sigma}_{ii}\rho - \rho\hat{\sigma}_{ii}]$$

Where $\Gamma_{4m}$ and $\Gamma_{5n}$ are the spontaneous emission rates out of state $|4\rangle$ and $|5\rangle$ to state $|m\rangle$ and $|n\rangle$ (m, n=1, 2, 3), respectively. The $\gamma_{ideph}$ is the dephasing rate of state $|i\rangle$ ($i=2, 3, 4, 5$), $\rho$ is the density matrix and $\sigma_{ij}$ is the projection operator. For convenience, we define the total decay rate out of state $|4\rangle$ and $|5\rangle$ are $\Gamma_4=\Gamma_{41}+\Gamma_{42}+\Gamma_{43}$, $\Gamma_5=\Gamma_{51}+\Gamma_{52}+\Gamma_{53}$ respectively. The coherence decay rates of these states are defined as: $\gamma_{21}=\gamma_{2deph}$, $\gamma_{31}=\gamma_{3deph}$, $\gamma_{32}=\gamma_{2deph}+\gamma_{3deph}$, $\gamma_{41}=\Gamma_4+\gamma_{4deph}$, $\gamma_{42}=\Gamma_4+\gamma_{4deph}+\gamma_{2deph}$, $\gamma_{43}=\Gamma_4+\gamma_{4deph}+\gamma_{3deph}$, $\gamma_{51}=\Gamma_5+\gamma_{5deph}$, $\gamma_{52}=\Gamma_4+\gamma_{5deph}+\gamma_{2deph}$ and $\gamma_{53}=\Gamma_5+\gamma_{5deph}+\gamma_{3deph}$. From equations (1) and (2) we derive the following optical Bloch equations for the off-diagonal density matrix elements in the steady-state condition:

$$\left[-\frac{\gamma_{21}i}{2} - (\Delta_1 - \Delta_s)\right]\rho_{21} - \frac{\Omega_p}{2}\rho_{25} + \frac{\Omega_{c1}^*}{2}\rho_{41} - \frac{\Omega_s}{2}\rho_{24} + \frac{\Omega_{as1}^*}{2}\rho_{51} = 0 \quad (3)$$

$$(-\frac{\gamma_{31}i}{2} + \Delta_s)\rho_{31} - \frac{\Omega_p}{2}\rho_{35} + \frac{\Omega_{c2}^*}{2}\rho_{41} - \frac{\Omega_s}{2}\rho_{34} + \frac{\Omega_{as2}^*}{2}\rho_{51} = 0 \quad (4)$$

$$(-\frac{\gamma_{32}i}{2}+\Delta_1)\rho_{32}+\frac{\Omega_{c2}^*}{2}\rho_{42}-\frac{\Omega_{c1}}{2}\rho_{34}+\frac{\Omega_{as2}^*}{2}\rho_{52}-\frac{\Omega_{as1}}{2}\rho_{35}=0 \tag{5}$$

$$(-\frac{\gamma_{41}i}{2}+\Delta_s)\rho_{41}+\frac{\Omega_{c1}}{2}\rho_{21}+\frac{\Omega_{c2}}{2}\rho_{31}+\frac{\Omega_s}{2}\rho_{11}=0 \tag{6}$$

$$(-\frac{\gamma_{42}i}{2}+\Delta_1)\rho_{42}+\frac{\Omega_{c1}}{2}\rho_{22}+\frac{\Omega_{c2}}{2}\rho_{32}+\frac{\Omega_s}{2}\rho_{12}=0 \tag{7}$$

$$-\frac{\gamma_{43}i}{2}\rho_{42}+\frac{\Omega_{c1}}{2}\rho_{23}+\frac{\Omega_{c2}}{2}\rho_{33}+\frac{\Omega_s}{2}\rho_{13}=0 \tag{8}$$

$$(-\frac{\gamma_{51}i}{2}+\Delta_p)\rho_{51}+\frac{\Omega_p}{2}\rho_{11}+\frac{\Omega_{as1}}{2}\rho_{21}+\frac{\Omega_{as2}}{2}\rho_{31}=0 \tag{9}$$

$$\left[-\frac{\gamma_{52}i}{2}+(\Delta_p+\Delta_1-\Delta_s)\right]\rho_{52}+\frac{\Omega_p}{2}\rho_{12}+\frac{\Omega_{as1}}{2}\rho_{22}+\frac{\Omega_{as2}}{2}\rho_{32}=0 \tag{10}$$

$$\left[-\frac{\gamma_{53}i}{2}+(\Delta_p-\Delta_s)\right]\rho_{53}+\frac{\Omega_p}{2}\rho_{13}+\frac{\Omega_{as1}}{2}\rho_{23}+\frac{\Omega_{as2}}{2}\rho_{33}=0 \tag{11}$$

$$\frac{\Omega_p}{2}\rho_{14}+\frac{\Omega_{as1}}{2}\rho_{24}+\frac{\Omega_{as2}}{2}\rho_{34}-\frac{\Omega_s^*}{2}\rho_{51}-\frac{\Omega_{c1}^*}{2}\rho_{52}-\frac{\Omega_{c2}^*}{2}\rho_{53}=0 \tag{12}$$

In the equations (3)-(12), we have assumed $\rho_{54}=0$. We consider the zeroth-order perturbation expansion with the assumption of $\Omega_{c1}$, $\Omega_{c2}$, $\Omega_p >> \Omega_s$, $\Omega_{as1}$, $\Omega_{as2}$ (our experimental conditions) and the initial atomic distribution in the ground state $|1>$ ($\rho_{11}=1$, $\rho_{22}=0$, $\rho_{33}=0$, $\rho_{44}=0$, $\rho_{55}=0$, initial state preparation), and derive the steady-state solutions of the above equations [12, 26, 27]. After some calculations, we obtain the first and third order nonlinear susceptibilities of this system, describing the generation of anti-Stokes fields ($\omega_{as1}$ and $\omega_{as2}$), respectively.

$$\rho_{52}^{(1)}=\frac{i\Omega_{as1}[\gamma_{21}+2i(\Delta_1-\Delta_s)]}{3\left\{[-\gamma_{52}-2i(\Delta_1+\Delta_p-\Delta_s)][\gamma_{21}+2i(\Delta_1-\Delta_s)]+|\Omega_p|^2\right\}} \tag{13}$$

$$\rho_{53}^{(1)}=\frac{i\Omega_{as2}(\gamma_{31}-2i\Delta_s)}{3\left\{[\gamma_{53}+2i(\Delta_p-\Delta_s)](\gamma_{31}-2i\Delta_s)+|\Omega_p|^2\right\}} \tag{14}$$

$$\rho_{52}^{(3)}=\frac{i\Omega_{c1}(\gamma_{32}-2i\Delta_1)\gamma_{43}\Omega_p\Omega_s^*}{3\left[(\gamma_{42}+2i\Delta_1)(\gamma_{32}\gamma_{43}-2i\gamma_{43}\Delta_1+|\Omega_{c1}|^2)+\gamma_{43}|\Omega_{c2}|^2\right]\left\{[-\gamma_{21}-2i(\Delta_1-\Delta_s)][\gamma_{52}+2i(\Delta_1+\Delta_p-\Delta_s)]+|\Omega_p|^2\right\}} \tag{15}$$

$$\rho_{53}^{(3)}=\frac{(\gamma_{32}+2i\Delta_1)(i\gamma_{42}+2\Delta_1)\Omega_{c2}\Omega_p\Omega_s^*}{3\left[(\gamma_{42}-2i\Delta_1)(\gamma_{32}\gamma_{43}+2i\gamma_{43}\Delta_1+|\Omega_{c1}|^2)+\gamma_{43}|\Omega_{c2}|^2\right]\left\{[\gamma_{53}+2i(\Delta_p-\Delta_s)](\gamma_{31}-2i\Delta_s)+|\Omega_p|^2\right\}} \tag{16}$$

The polarization in the atomic medium generated by the applied field is defined as $P=Nu\rho$ (where, $N$ is the density of atoms, $u$ is the dipole moment and). The polarization plays the part of the source term in Maxwell's equations and determines the electromagnetic field dynamics. In our system, the powers of generated anti-Stokes fields ($\omega_{as1}$ and $\omega_{as2}$) are very small with respect to the pump field. The amplitude of the FWM signal fields we calculated are proportional to the third-order nonlinear susceptibilities $\rho_{52}^{(3)}$ and $\rho_{53}^{(3)}$, We could use these expressions to simulate our experimental results.

## 3. Experimental setup and results

We now turn to the experimental part of this letter. A two-dimensional magneto-optical trap of $^{85}$Rb with a maximal optical depth (OD) of 38 is employed in this experiment [25]. We use an amplified external cavity diode laser (ECDL) (TA100, 780nm, Toptica) as the cooling beams and another ECDL (DL100, 780nm, Toptica) as the repumper. The TA100 has an output power of 430mW, locked to 15MHz to the red of the $5S_{1/2}$, (F=3)—>$5P_{3/2}$, (F=4) transition. After beam modulating and single-mode fiber filtering, we have 100mW cooling power available in the vacuum chamber. The repumper with an available power of 25mW is locked resonantly to the $5S_{1/2}$, (F=2)—>$5P_{3/2}$, (F=2) transition.

All the four beams are collimated as Fig.2 so as to satisfy the phase-matching conditions simultaneously. The probe beam ($\omega_s$) injects along the long axis of the cigar-shaped atomic cloud and is focused to the center of the cloud. The pump beam ($\omega_p$) with σ+ polarization counter propagates with the two coupling beams ($\omega_{c1}$ and $\omega_{c2}$, σ-polarized) and all the three beams overlap the cigar-shaped atomic cloud. The coupling-pump axis (coupling and pump beams) has an angle of 2 degrees from the Stokes-anti-Stokes axis (probe beam and FWM signals). All the lasers used in the experiment are provided by three commercial ECDLs (DL100, Toptica), and pass through the respective acousto-optic modulators (AOMs) for beam switching and frequency shifting. The transmission spectrum of the probe beam is detected by a photomultiplier tube PMT1 (H10721, Hamamatsu). And we use PMT2 to detect the generated FWM signals ($\omega_{as1}$ and $\omega_{as2}$).

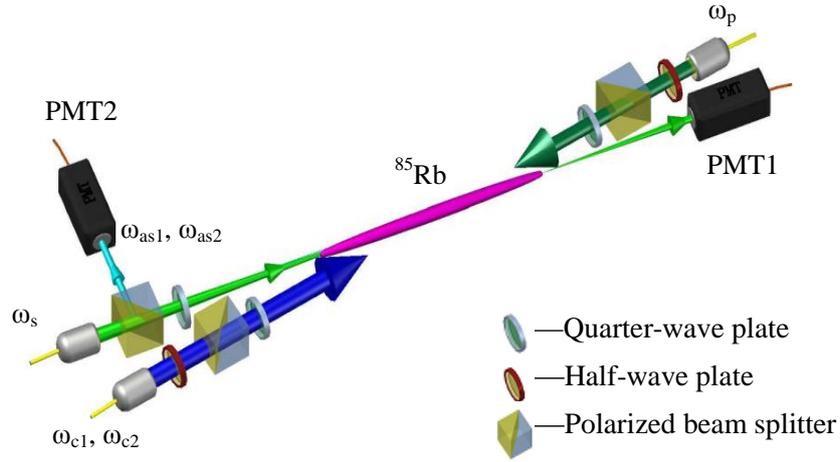

**Figure 2.** (color online) Experimental configuration for the double-EIT and double-FWM. The angle between Stokes-anti-Stokes axis and coupling-pump axis is 2°. This angle can filter noise from the pump and coupling beams. The Stokes beam is focused to the center of the atom cloud, and pump and coupling beams are collimated so as to totally overlap the Stokes beam in the cloud.

The atoms are initially prepared on the state $|5S_{1/2}, F=3, m_F=-3\rangle$. By turning off the MOT trapping laser while keeping the repumper on for 200 μs, the atoms are prepared at the ground level $|1\rangle$, as shown in Fig.1. The experiment is performed in a 700 μs window after shutting off the trapping and repumping beams. The whole process repeats every 33 ms. We lock the coupling beams $\omega_{c1}$ and $\omega_{c2}$ in different frequency detunings near the respective transitions and scan the probe field frequency in the detuning range of -40~40MHz near the $|5^2S_{1/2}, F=3\rangle \rightarrow |5^2P_{1/2}, F=3\rangle$ transition. Here, the pump beam is blocked. The two coupling beams ($\omega_{c1}$ and $\omega_{c2}$) will create transparent windows simultaneously for the probe field ($\omega_s$) (see Fig.1). Depending on the frequency detunings of the two coupling beams, these two EIT windows can either overlap or be separated in frequency on the probe beam transmission signal(as shown in Fig.3). We could find the two transparent widows for the probe field in the three curves of Fig.3.

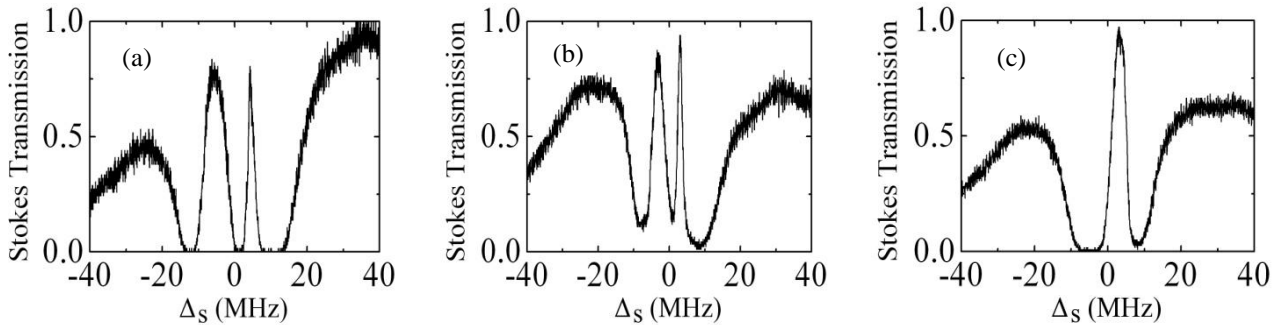

**Figure 3.** Observation of double-EIT versus the probe detuning. The different width of the two transparent windows is due to the different Rabi frequencies of coupling beams $\omega_{c1}$ and $\omega_{c2}$. The coupling beams $\omega_{c1}$ is locked

5MHz to the red of the respective transition, and we tune the detuning of $\omega_{c2}$ (-5MHz in (a) and -3MHz in (b)). The two coupling beams have the same detuning in (c).

We then inject a pump beam on the atomic cloud and build up the FWM configuration. There are two possible phase matching conditions $\vec{k}_p + \vec{k}_{c1} + \vec{k}_s + \vec{k}_{as1} = 0$ and $\vec{k}_p + \vec{k}_{c2} + \vec{k}_s + \vec{k}_{as2} = 0$. The simultaneously opened double-EIT windows in this five-level atomic system allow observation of these two nonlinear optical processes at the same time, both of the two phase-matching conditions could be satisfied simultaneously. The results are shown in Fig. 4. By individually controlling (or tuning) the EIT windows, the generated two FWM signals can be clearly separated and distinguished or pulled together (by adjusting the coupling frequency detunings). The generated two anti-Stokes fields have the frequencies of $\omega_{as1} = \omega_p + \omega_{c1} - \omega_s$ and $\omega_{as2} = \omega_p + \omega_{c2} - \omega_s$, respectively. We tune the frequency detuning of the coupling beam $\omega_{c1}$ from -21.5MHz to 16MHz and lock the coupling beam $\omega_{c2}$ resonant to the respective transition. We then similarly scan the probe frequency. It is found that, not only the transparent windows for the probe field but also the generated anti-Stokes field $\omega_{as1}$ will follow the coupling beam ($\omega_{c1}$) detuning. The pump beam will induce transparent windows for both the two anti-Stokes fields because of EIT effect. The full widths at half maximum (FWHM) of the two FWM peaks, which are measured to be 6.7MHz for $\omega_{as1}$ and 2.3MHz for $\omega_{as2}$, are determined by the Rabi frequencies of the pump and respective coupling beams.

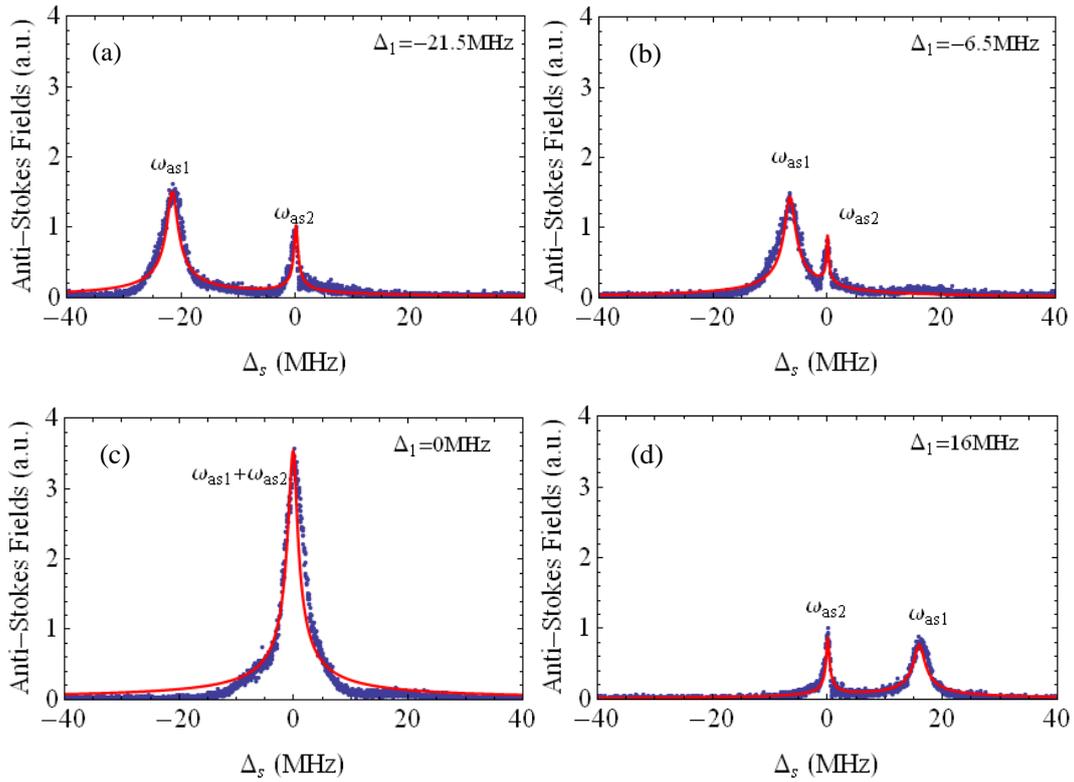

**Figure 4**. Observation of double peaks in the FWM spectrum versus the probe frequency detuning under different coupling frequency detunings. The experimental results (blue dot) fit well with the theoretical simulation (red line). The simulating parameters are: $\gamma=2\pi\times 6$MHz, $\Omega_s=2\gamma$, $\Delta_p=25\gamma$, $\Delta_s=0$, $\Omega_{c1}=3.6\gamma$, $\Omega_{c2}=2.4\gamma$, $\Omega_p=4.0\gamma$, $\gamma_{2deph}=0.02\gamma$, $\gamma_{3deph}=0.008\gamma$, OD=38.

We could see from Fig.4 that the two different types of FWM processes could be realized simultaneously, which means the phase matching and energy conservation could be satisfied for both of the two processes in this experimental configuration. As being shown in Fig.4(c), if the coupling beams $\omega_{c1}$ and $\omega_{c2}$ have the same frequency detuning to the respective transitions, the generated anti-Stokes signals will overlap with a frequency difference which is equal to the ground state hyperfine splitting (3GHz). Thus a bichromatic field could be carried out from this five-level system. It is indicated in Fig. 4 that the theoretically red curves in the four panels, which come from the third-order nonlinear susceptibilities $\rho_{52}^{(3)}$ and $\rho_{53}^{(3)}$, agree well with the experimental data lines (blue). In our system, the powers of generated anti-Stocks fields ($\omega_{as1}$, $\omega_{as2}$) are very small with respect to the pump and coupling fields. So we can ignore the linear susceptibilities. The parameters used in the simulation are: $\gamma=2\pi\times6MHz$, $\Omega_s=2\gamma$, $\Delta_p=25\gamma$, $\Delta_s=0$, $\Omega_{c1}=3.6\gamma$, $\Omega_{c2}=2.4\gamma$, $\Omega_p=4.0\gamma$, $\gamma_{2deph}=0.02\gamma$, $\gamma_{3deph}=0.008\gamma$, OD=38.

## 4. Conclusion

In conclusion, we have experimentally demonstrated and theoretically modeled the bichromatic field generation from a double-FWM system. The theoretical simulation agrees well with the experimental results (see Fig.4). The bichromatic field generated from the two FWM processes could be a potential candidate for the color-entanglement and cross phase modulation. The two colors in the bichromatic field could have the same polarization, intensity and spatial mode except for the color. Such coexistence of two FWM processes allows us to investigate the interplay between these two nonlinear optical effects. If we assume that the light-atom interaction time is short or the intensities of the coupling and pump fields are weak, both the mean photon numbers in the two generated FWM pulses are much smaller than 1. Thus the photon state at the low light level could be seem as the superposition of the two frequencies $|\omega_{as1}\rangle+|\omega_{as2}\rangle$. What's more, adjusting one of the FWM signals will have effect on the other. The group velocities of these two FWM fields could be adjusted by tuning the intensity or frequency detuning of the pump beam. Such a bichromatic field could also be useful in the cross phase modulation of weak pulses at the low light level.

## Acknowledgements

We would like to thank Professor Jianming Wen for his helpful discussion. This work was supported by the National Natural Science Foundation of China (Grant Nos. 10874171, 11174271) and the National Fundamental Research Program of China (Grant No. 2011CB00200).

## References

[1] M. Fleischhauer, A. Imamoglu, and J. P. Marangos 2005 *Rev. Mod. Phys.* **77** 633

[2] Harris S E 1997 *Phys. Today* **50** 36

[3] Boller K J, Imamoglu A and Harris S E 1991 *Phys. Rev. Lett.* **66** 2593

[4] Li Y. and Xiao M. 1996 *Opt. Lett.* **21** 1064; Lu B., Burkett W. H. and Xiao M. 1998 *Opt. Lett.* **23** 804

[5] Zibrov A. S., Matsko A. B., Kocharovskaya O., Rostovtsev Y. V., Welch G. R. and Scully M. O. 2002 *Phys. Rev. Lett.* **88** 103601

[6] Rostovtsev Y. V., Sariyanni Z. E. and Scully M. O. 2006 *Phys. Rev. Lett.* **97** 113001

[7] Balic V., Braje D. A., Kolchin P., Yin G. Y. and Harris S. E. 2005 *Phys. Rev. Lett.* **94** 183601

[8] D. F. Phillips, A. Fleischhauer, A. Mair, R. L. Walsworth and M. D. Lukin 2001 *Phys. Rev. Lett.* **86**, 783–786

[9] N. B. Phillips, A. V. Gorshkov and I. Novikova 2011 *Phys. Rev.* A **83** 063823

[10] Y. -F. Chen, S. -H. Wang, C. -Y. Wang, and I. A.Yu 2005 *Phys. Rev.* A **72** 053803

[11] Y. -F. Chen, P. -C. Kuan, S. -H. Wang, C. -Y. Wang, and I. A. Yu 2006 *Opt. Lett.* **31** 3511–3513

[12] Shengwang Du, Jianming Wen and Morton H. Rubin 2008 *J. Opt. Soc. Am.* B **25** C98

[13] Shengwang Du, Pavel Kolchin, Chinmay Belthangady, G.Y. Yin and S.E. Harris 2008 *Phys. Rev. Lett.* **100** 183603

[14] Pavel Kolchin, Shengwang Du, Chinmay Belthangady, G. Y. Yin and S. E. Harris 2006 *Phys. Rev. Lett.* **97** 113602

[15] V. Balic' et al. 2005 *Phys. Rev. Lett.* **94** 183601

[16] Shiau, B.-W., Wu, M.-C. Lin, C.-C. and Chen, Y.-C. 2011 *Phys. Rev. Lett.* **106** 193006

[17] H. Schmidt and A. Imamog lu 1996 *Opt. Lett.* **21** 1936

[18] L.V. Hau et al. 1999 *Nature* (London) **397** 594

[19] H. Kang and Y. Zhu 2003 *Phys. Rev. Lett.* **91** 093601

[20] S. Ramelow et al. 2009 *Phys. Rev. Lett.* **103** 253601

[21] Y. P. Zhang, A. W. Brown, and M. Xiao 2007 *Phys.Rev.Lett.* **99** 123603

[22] Y. P. Zhang, U. Khadka, B. Anderson, and M. Xiao 2009 *Phys. Rev. Lett.* **102** 013601

[23] X.H. Yang, J.T. Sheng, U. Khadka and M. Xiao 2011 *Phys. Rev.* A **83** 063812

[24] X.H. Yang, J.T. Sheng, U. Khadka and M. Xiao 2012 *Phys. Rev.* A **85** 013824

[25] Y. Liu, J.H. Wu, B.S. Shi and G.C. Guo 2012 *Chin. Phys. Lett.* **29** 024205

[26] J.-M. Wen and M. H. Rubin 2006 *Phys. Rev.* A **74** 023808

[27] J.-M. Wen and M. H. Rubin 2006 *Phys. Rev.* A **74** 023809

[28] J. Wang, Y. Zhu, K. J. Jiang and M. S. Zhan 2003 *Phys.Rev.* A **68** 063810

[29] G. Q. Yang, P. X u, J. Wang, Yifu Zhu and M. S. Zhan 2010 *Phys.Rev.* A **82** 045804

[30] Y. Wu 2005 *Phys.Rev.* A **71** 053820

[31] Y. Wu and X. Yang 2004 *Phys.Rev.* A **70** 053818